\documentclass[prl, reprint, superscriptaddress]{revtex4-1}
\usepackage{amsmath}
\usepackage{graphicx}
\usepackage{graphicx}
\usepackage{amsmath}
\usepackage{amssymb}
\usepackage{url}
\usepackage{bm}		
\usepackage{color}	
\usepackage{textcomp}	

\begin{document}	

\title{Doubly-resonant $\chi^{(2)}$ nonlinear photonic crystal cavity based on a bound state in the continuum}

\author{Momchil Minkov}
\email[]{mminkov@stanford.edu}
\affiliation{Department of Electrical Engineering, and Ginzton Laboratory, Stanford University, Stanford, CA 94305, USA}

\author{Dario Gerace}
\affiliation{Dipartimento di Fisica, Universit\`a di Pavia, via Bassi 6, 27100 Pavia, Italy}

\author{Shanhui Fan}
\email[]{shanhui@stanford.edu}
\affiliation{Department of Electrical Engineering, and Ginzton Laboratory, Stanford University, Stanford, CA 94305, USA}

\begin{abstract}
Photonic nanostructures simultaneously maximizing spectral and spatial overlap between fundamental and second-harmonic confined modes are highly desirable for enhancing second-order nonlinear effects in nonlinear media. These conditions have thus far remained challenging to satisfy in photonic crystal cavities because of the difficulty in designing a band gap at the second-harmonic frequency. Here, we solve this issue by using instead a bound state in the continuum at that frequency, and we design a doubly-resonant photonic crystal slab cavity with strongly improved figures of merit for nonlinear frequency conversion when compared to previous photonic crystal designs. Furthermore, we show that the far-field emission at both frequencies is highly collimated around normal incidence, which allows for simultaneously efficient pump excitation and collection of the generated nonlinear signal. Our results could lead to unprecedented conversion efficiencies in both parametric down conversion and second harmonic generation in an extremely compact architecture.
\end{abstract}

\maketitle

\section{Introduction}

Highly efficient second-order nonlinear frequency conversion would heavily impact a number of applications in current nanophotonics research, such as biosensing \cite{Freund1986, Campagnola2003}, nonlinear spectroscopy \cite{Heinz1982, Rakher2010}, and efficient nonclassical radiation sources \cite{Caspani2017}. For these purposes, developing compact devices with high conversion efficiency is of significant interest, and there has been substantial effort towards designing resonant cavity structures with the aim of enhancing the efficiency of nonlinear frequency conversion. 

For second-order nonlinear materials, the cavities that offer the strongest enhancement of the frequency conversion are doubly-resonant, i.e. supporting resonances simultaneously at both the first harmonic (FH) and the second harmonic (SH) frequencies \cite{Berger1997, Liscidini2006, Scaccabarozzi2006, Hayat2007}. Achieving such doubly-resonant cavities for nonlinear frequency conversion represents a long-standing challenge in optical device design. In addition to the stringent requirement of the double resonance, the nonlinear overlap factor, involving the fields of the first and the second harmonic modes and the spatial distribution of the $\chi^{(2)}$ tensor,  must be sufficiently large to ensure an efficient frequency conversion process. Doubly-resonant cavities have been demonstrated in on-chip ring-resonator structures \cite{Pernice2012, Bruch2018}, but with relatively large footprint. Recently, further progress was made by exploiting topology optimization to design small-footprint micropost and microring cavities with enhanced theoretical nonlinear efficiency \cite{Lin2016, Lin2017}. Due to the topology optimization, however, these devices incorporate a number of very fine structural features which make them challenging to fabricate in practice.

Photonic crystal (PhC) defect cavities patterned in thin slabs support ultra-long lived resonant modes confined in volumes approaching the diffraction limit \citep{Akahane2003, Song2005}, which makes them attractive candidates for a number of applications, including nonlinear harmonic generation. In addition, they also provide a high degree of far-field radiation pattern engineering \cite{Tran2009, Portalupi2010}, which could facilitate in-coupling from and out-coupling to free space. Singly-resonant photonic crystal slab cavities patterned in nonlinear materials have already shown very large nonlinear conversion efficiencies \cite{Galli2010, Buckley2014a, Yamada2014, Zeng2015, Mohamed2017}, despite the lack of a confined resonance at the second harmonic. This promises that engineering doubly resonant modes in such structures could yield unprecedented enhancement. However, designing confined modes fulfilling the doubly resonant condition with either one-dimensional nanobeams or two-dimensional photonic crystal slab cavities has proven very challenging \cite{Rivoire2009, Rivoire2011, Buckley2014, Andreani2006, Minkov2014a}. Photonic crystal slab cavities are usually designed starting from a partial photonic band gap in the guided mode spectrum. This strategy can be successfully applied to design a high-$Q$ cavity at the first harmonic. However, the second harmonic frequency range generally lies entirely inside the light cone of the slab, such that not even a partial photonic band gap is possible. As a result, in spite of several decades of efforts in optimizing photonic crystal cavity designs, there has not been any report of a successful doubly-resonant photonic crystal cavity structure for second-order nonlinear frequency conversion. 

\begin{figure}
\centering
\includegraphics[width = \columnwidth]{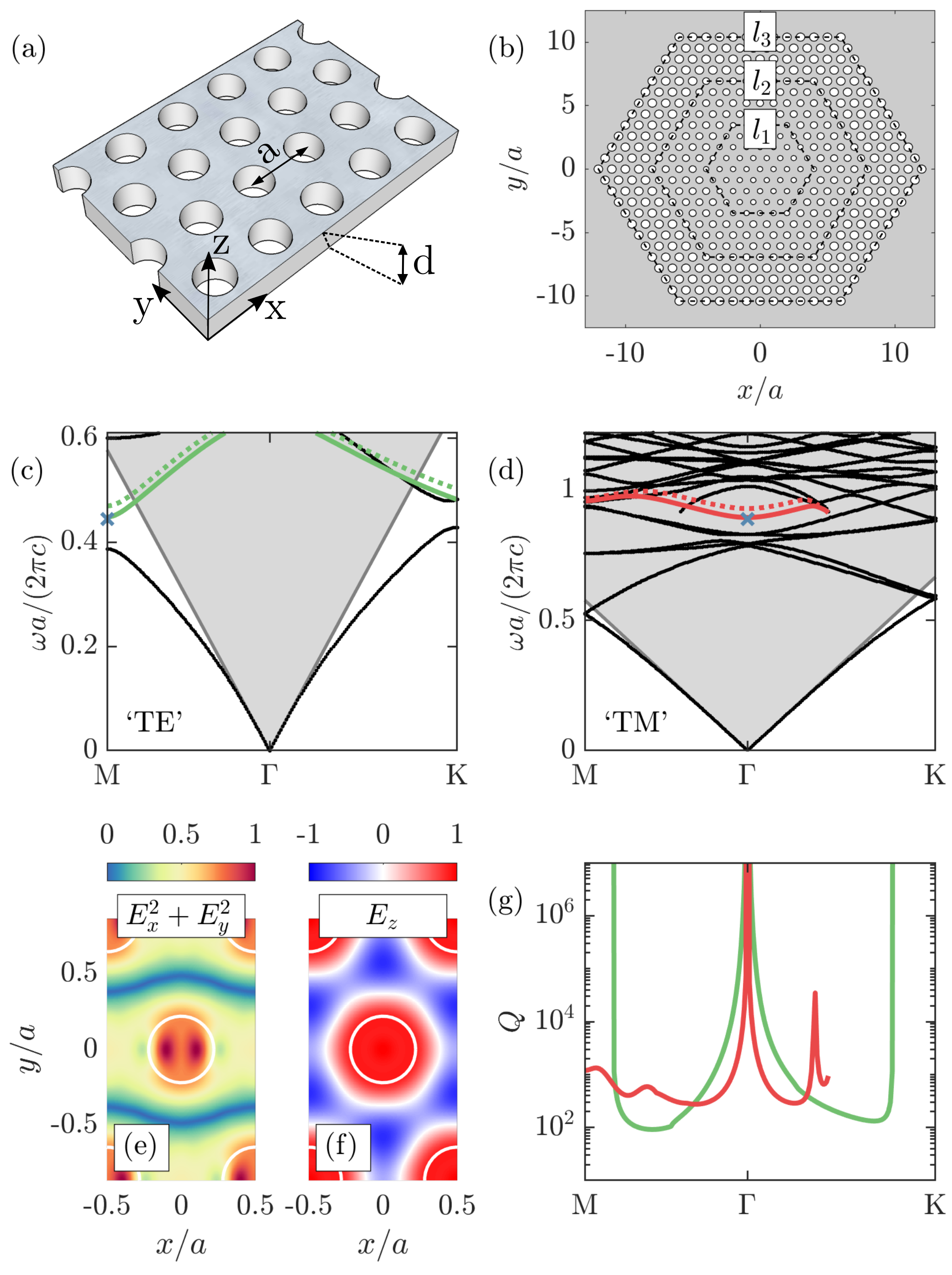}%
 \caption{(a): Schematic of a photonic crystal slab of thickness $d$ with a hexagonal lattice with lattice period $a$ of air holes with radius $r$. (b): Schematic of a heterostructure cavity -- dashed hexagons show the core, transition, and outer regions, which have $N_c$, $N_t$, and $N_o$ layers of holes, respectively, such that the hexagon side-lengths are $l_1 = N_c a$, $l_2 = (N_c + N_t)a$, and $l_3 = (N_c + N_t + N_o)a$. In the image, $N_c = N_t = N_o = 4$, and the corresponding hole radii are $r_c = 0.2a$, $r_t = 0.26a$, and $r_o = 0.32a$. (c): Photonic bands for the quasi-TE modes for a regular PhC slab with $d = 0.28a$, $r = 0.22a$, and refractive index $n = 2.32$. The light cone is shaded gray. (d): Photonic bands for the quasi-TM modes for the same $d$ and $r$ as in (c), and slab refractive index $n = 2.38$. The green and red bands in panels (c)-(d) highlight the bands from which the heterostructure modes are later derived. The dotted green/red lines show the corresponding bands computed for $r = 0.25a$. (e): Electric field $E_x^2 + E_y^2$ the center of the slab for the mode at the $M$-point marked by a cross in (c), with a $\mathbf{k}$-vector in the $y$-direction. (f): Electric field $E_z$ in the center of the slab for the mode at the $\Gamma$-point marked by a cross in (d). (g): Radiative quality factors for the two highlighted bands in (c)-(d).}
\label{fig:phc_def}
\end{figure} 

In this paper, we introduce a design of a doubly resonant photonic crystal cavity with high modal overlap factor for $\chi^{(2)}$ nonlinear frequency conversion. This design is built upon the concept of bound states in the continuum (BIC), which has recently received renewed interest in photonics \cite{Hsu2016}. BICs can be observed in various systems, but in PhC slabs in particular they correspond to modes that lie inside the light cone but are nonetheless non-radiative, fully-guided inside the slab. Thus, the key idea here lies in abandoning the commonly-held notion that having photonic band gaps at both the first and second harmonics is an essential prerequisite to design doubly-resonant PhC cavities. Indeed, it was recently shown that a heterostructure cavity can be realized in a two-dimensional PhC slab without a band gap \cite{Ge2018}. The concept is illustrated in Fig. \ref{fig:phc_def}(a)-(b). The starting structure is a regular PhC made by a hexagonal lattice with lattice period $a$ of air holes of radius $r$ in a slab with thickness $d$ and refractive index $n$ (Fig. \ref{fig:phc_def}(a)). A heterostructure design can then be formed by introducing PhC regions with different hole radii, as shown in Fig. \ref{fig:phc_def}(b). Even with no band gap, modes localized in the core region of such a structure can still be observed close to a band-edge frequency of the photonic bands of the underlying PhC. When these band-edge modes have a diverging quality factor due to a BIC effect, a large $Q$ can also be expected for the heterostructure modes -- in particular in the limit of increasing core size \cite{Ge2018}. In our previous work \cite{Ge2018}, we also found that having a transition region between the core and the outer regions (Fig. \ref{fig:phc_def}(b)) further improves the quality factor of the heterostructure cavity.

Below, we use this route to design a high-$Q$ localized mode at the SH frequency, and to precisely tailor the doubly resonant condition, starting from photonic bands that fulfill the right symmetry properties for efficient $\chi^{(2)}$ nonlinear frequency conversion. To show the results on a practical photonic platform, we explicitly choose the slab material to be Gallium Nitride (GaN), in which efficient continuous wave SH generation has been observed for singly resonant photonic crystal slab cavities with FH in the near infrared \cite{Zeng2015, Mohamed2017}. The design strategy hereby introduced could be extended to different wavelength ranges as well as other material platforms. In addition to the applications in frequency conversion, these results could also prove important towards exploiting second-order nonlinearities to reach the regime of photon blockade \cite{Majumdar2013, Gerace2014}, thus opening the door to the engineering of single-photon Fock states in integrated quantum photonics.

\section{Recipe for band engineering}

We begin our study by designing the photonic bands of the GaN photonic crystal slab as shown in Fig. \ref{fig:phc_def}(a) to satisfy the following conditions:
\begin{enumerate}
\item A band with a band edge mode outside the light cone at the FH frequency $\omega_1$.
\item A band with a singly-degenerate band edge mode at the $\Gamma$-point at a frequency $\omega_2 \approx 2\omega_1$.
\item The band-edge modes in 1. and 2. are simultaneously either a maximum or a minimum of their corresponding bands.
\item The periodic parts of the electric field of these two band-edge modes have a nonzero nonlinear overlap factor.
\end{enumerate}

Below we discuss the rationales behind these conditions. The nonlinear overlap factor between two resonant modes for second harmonic generation is defined as \cite{Rodriguez2007}
\begin{equation}
\beta_1 = \frac{1}{4} \frac{\int \mathrm{d} \mathbf{r} \epsilon_0 \sum_{ijk}\chi^{(2)}_{ijk}(\mathbf{r}) \left(E^*_{1i} E_{2j} E^*_{1k} + E^*_{1i} E^*_{1j} E_{2k} \right)}{\left(\int \mathrm{d}\mathbf{r} \epsilon_0 \epsilon_1(\mathbf{r}) |\mathbf{E}_1|^2 \right)\left(\int \mathrm{d}\mathbf{r} \epsilon_0 \epsilon_2(\mathbf{r}) |\mathbf{E}_2|^2 \right)^{1/2}},
\label{eqn:beta}
\end{equation}
where $\chi^{(2)}_{ijk}$ is the second-order nonlinear susceptibility tensor for the material, $\mathbf{E}_1$ and $\mathbf{E}_2$ are the mode profiles and $\epsilon_1(\mathbf{r})$ and $\epsilon_2(\mathbf{r})$ are the relative permittivity profiles at the FH and the SH frequencies, respectively. Here, we take into account that the permittivities in general are different for the first and second harmonic modes. Assuming the GaN slab is grown with its c-axis along the $z$-direction (orthogonal to the slab plane), the dominant $\chi^{(2)}$ components are $xxz$, $yyz$ and $zzz$ \cite{Sanford2005}. In a PhC that is symmetric with respect to the $xy$-plane bisecting the slab, the positive- and negative-symmetry photonic modes are predominantly $E_{x,y}$ and $E_z$ polarized, respectively (quasi-TE and quasi-TM). Thus, the two modes in Conditions 1 and 2 can be either both quasi-TM (coupling through $\chi^{(2)}_{zzz}$), or quasi-TE at the FH and quasi-TM at the SH (coupling through $\chi^{(2)}_{xxz}$ and $\chi^{(2)}_{yyz}$). Since the hexagonal lattice PhC slab has a photonic band gap for the quasi-TE modes at low frequencies, which could help with the localization and the quality factor of the heterostructure modes, we will look for a quasi-TE FH and a quasi-TM SH mode. Of course, for other materials or other orientations of the GaN slab, the relevant $\chi^{(2)}$ components have to be considered when making this choice.

In the end of this section, we will also discuss how Conditions 1 and 2 ensure not only the resonance matching ($\omega_2 \approx 2\omega_1$), but also that the heterostructure modes are derived from a photonic crystal mode with an infinite quality factor, and hence can have high $Q$-s at both the FH and the SH frequencies. Condition 3 is simply required to ensure that a heterostructure cavity will localize light in its central region for both frequencies, and needs no further discussion. Condition 4 imposes mode-matching for the periodic parts of the Bloch modes, and will be further discussed below.

We use the guided-mode expansion method \cite{Andreani2006} to scan the parameters $d$ and $r$ for the photonic crystal in order to find bands that match all four conditions above. The material dispersion is taken into account by setting two different values of the slab refractive index for the quasi-TE and the quasi-TM simulations. Namely, we set $n = 2.32$ and $n = 2.38$, as for GaN at wavelengths around $1.3$\textmu m and $0.65$\textmu m, respectively. In Fig. \ref{fig:phc_def}(c), we show the photonic band structure for the quasi-TE modes of the PhC with $d = 0.28a$, $r = 0.22a$ and $n = 2.32$, while in panel (d) we show the quasi-TM modes of the same PhC but with $n = 2.38$. As can be seen, there is a pair of bands that satisfy conditions 1, 2 and 3 from our list, highlighted in green in panel (c), and in red in panel (d). 

Beyond the polarization matching, it is also crucial to consider the fact that $\beta_1$ is proportional to a spatial integral over the local field contributions that can add up constructively as well as destructively. This generally imposes phase-matching conditions for guided modes, while, for localized modes where the average $\mathbf{k}$-vector is zero, the mode profiles must still be engineered such that $\beta_1$ does not vanish. In other words, even for modes that are co-localized in the same region of space, we must also ensure that the integral in the numerator of eq. (\ref{eqn:beta}) is large. For a heterostructure cavity, a localized mode derived from a single band is expected to be well approximated by
\begin{equation}
\mathbf{E}_{n,p}(\mathbf{r}) = \int_{BZ} \mathrm{d}\mathbf{k} f_{n,p}(\mathbf{k}) e^{i\mathbf{k}\cdot\bm{\rho}} \mathbf{E}_{\mathbf{k}n}(\mathbf{r}) ,
\label{eqn:E_het}
\end{equation}
where $\bm{\rho}$ is the in-plane component of $\mathbf{r}$, $\mathbf{E}_{\mathbf{k}n} (\mathbf{r})e^{i\mathbf{k}\cdot\bm{\rho}}$ are the Bloch modes of the $n$-th band from which the heterostructure mode is derived, and $f_{n,p}(\mathbf{k})$ is some envelope function. The index $p$ reflects the fact that a number of envelope functions could be supported. Assuming that $f_{n, p}(\mathbf{k})$ is narrowly centered around the band-edge, as must be the case in the limit of a heterostructure size spanning multiple lattice periods, we can approximate $\mathbf{E}_{\mathbf{k}n}(\mathbf{r}) \approx \mathbf{E}_{\mathbf{k}_0n}(\mathbf{r})$, where $\mathbf{k}_0$ is the band-edge Bloch vector. Note that, due to the $C_6$ and time-reversal symmetries, $\mathbf{E}_{\mathbf{k}n}(\mathbf{r})$ is the same for any of the possible $\mathbf{k}_0$ choices for a band edge at the M or the K points, and must also be real-valued.

Given these considerations, in Supplement 1 we study the overlap terms that need to be non-vanishing for FH and SH modes of the form of eq. (\ref{eqn:E_het}). One of the terms is between the envelope functions $f (\mathbf{k})$, and we find it to be always non-zero when the SH envelope $f_{2, p'} (\mathbf{k})$ is centered around $\Gamma$. Intuitively, this result stems from the fact that, in a homogeneous material, an SH guided mode at $\Gamma$ (i.e. constant $\mathbf{E}$-field in real space) is phase-matched with an FH standing-wave field of the form $\cos(\mathbf{k}\rho)$ for any $\mathbf{k}$. This is because the FH field would enter the overlap term in eq. (\ref{eqn:beta}) as $\cos^2(\mathbf{k}\rho)$. The second overlap integral that must be non-vanishing is between $\mathbf{E}_{\mathbf{k}_0 1}(\mathbf{r})$ and $\mathbf{E}_{\mathbf{k}_0' 2}(\mathbf{r})$, i.e. the periodic parts of the PhC Bloch modes at the FH and SH frequency. Looking at eq. (\ref{eqn:beta}), we note that, regardless of the symmetries of the $\mathbf{E}_{\mathbf{k}_0 1}$ profile, the spatial dependence of $E_{\mathbf{k}_01x}^2$ and $E_{\mathbf{k}_01y}^2$ is always even with respect to reflections both in the $xz$ and in the $yz$ planes, as can is also evident in Fig. \ref{fig:phc_def}(e). Thus, condition 4 translates to the requirement that the $E_{\mathbf{k}_0' 2z}$ profile must also be even with respect to both reflections. We took this into account in designing our PhC bands, as can be seen in Fig. \ref{fig:phc_def}(f).

\begin{figure*}
\centering
\includegraphics[width = 0.9\textwidth]{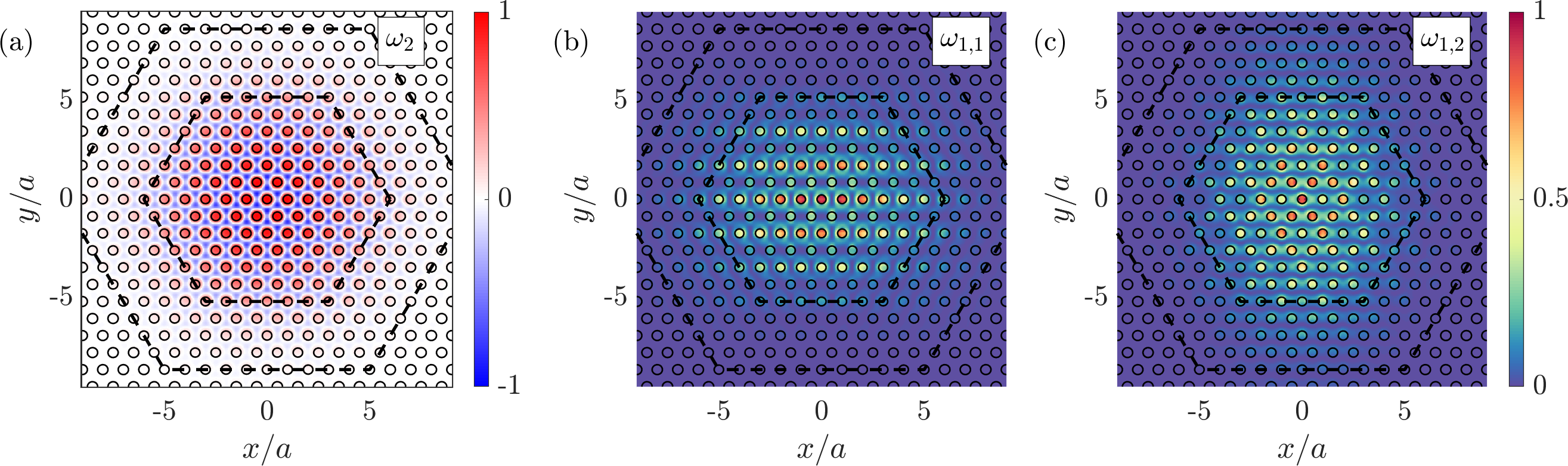}%
 \caption{(a): Electric field $\mathcal{R}(E_z)$ in the center of the slab for a heterostructure cavity mode at reduced frequency $\bar{\omega}_2 = \omega_2 a/(2\pi c) = 0.889$ in a PhC with $d = 0.28a$. The heterostructure parameters are $(N_c, N_t, N_o) = (6, 4, 14)$ and $(r_c, r_t, r_o) = (0.22a, 0.23a, 0.25a)$, and the dashed hexagons mark the core and transition regions. (b)-(c): Electric field $|E_x^2 + E_y^2|$ in the slab center for the same cavity, for two degenerate resonant modes at frequency $\bar{\omega}_1 = 0.444$. The slab refractive index was set to $2.38$ in (a) and $2.32$ in (b) and (c). For a more detailed image of the structure and the calculated modes, see Supplement 1.}
\label{fig:het_cavity}
\end{figure*} 

Finally, since the mode in a heterostructure cavity is derived from the band-edge modes in a photonic crystal as shown in the discussions below Eq. (2), to form a high-$Q$ heterostructure cavity it is desirable that the corresponding band-edge mode has infinite $Q$. We note that there are two ways to achieve an infinite $Q$: first, band modes outside the light cone of a PhC slab (gray shaded region in Fig. \ref{fig:phc_def}(c)-(d)) have infinite intrinsic lifetime and are perfectly confined around the slab region \cite{Johnson1999}. In contrast, inside the light cone there is a continuum of radiation modes, as well as a number of quasi-guided bands that are partially confined in the slab, but most generally have a finite radiative lifetime. Even in that region, however, it is possible to have non-radiative bound states in the continuum, either due to symmetry or due to an accidental destructive interference of the radiated light \cite{Hsu2016}. In particular, any singly-degenerate mode at the $\Gamma$ point of the band structure must necessarily be non-radiative due to symmetry \cite{Ochiai2001, Fan2002}. This is what motivates our particular requirement in Condition 2 for a singly-degenerate mode at $\Gamma$. To confirm our expectations, in Fig. \ref{fig:phc_def}(g), we plot the radiative quality factors corresponding to the two bands highlighted in Fig. \ref{fig:phc_def}(c)-(d). As can be seen, the $Q$ of the green band at the band minimum at the $M$-point goes to infinity since this mode lies outside the light cone, while the $Q$ of the red band at the band minimum at $\Gamma$ goes to infinity since the mode is a BIC. 

\begin{figure}[h!]
\centering
\includegraphics[width = \columnwidth]{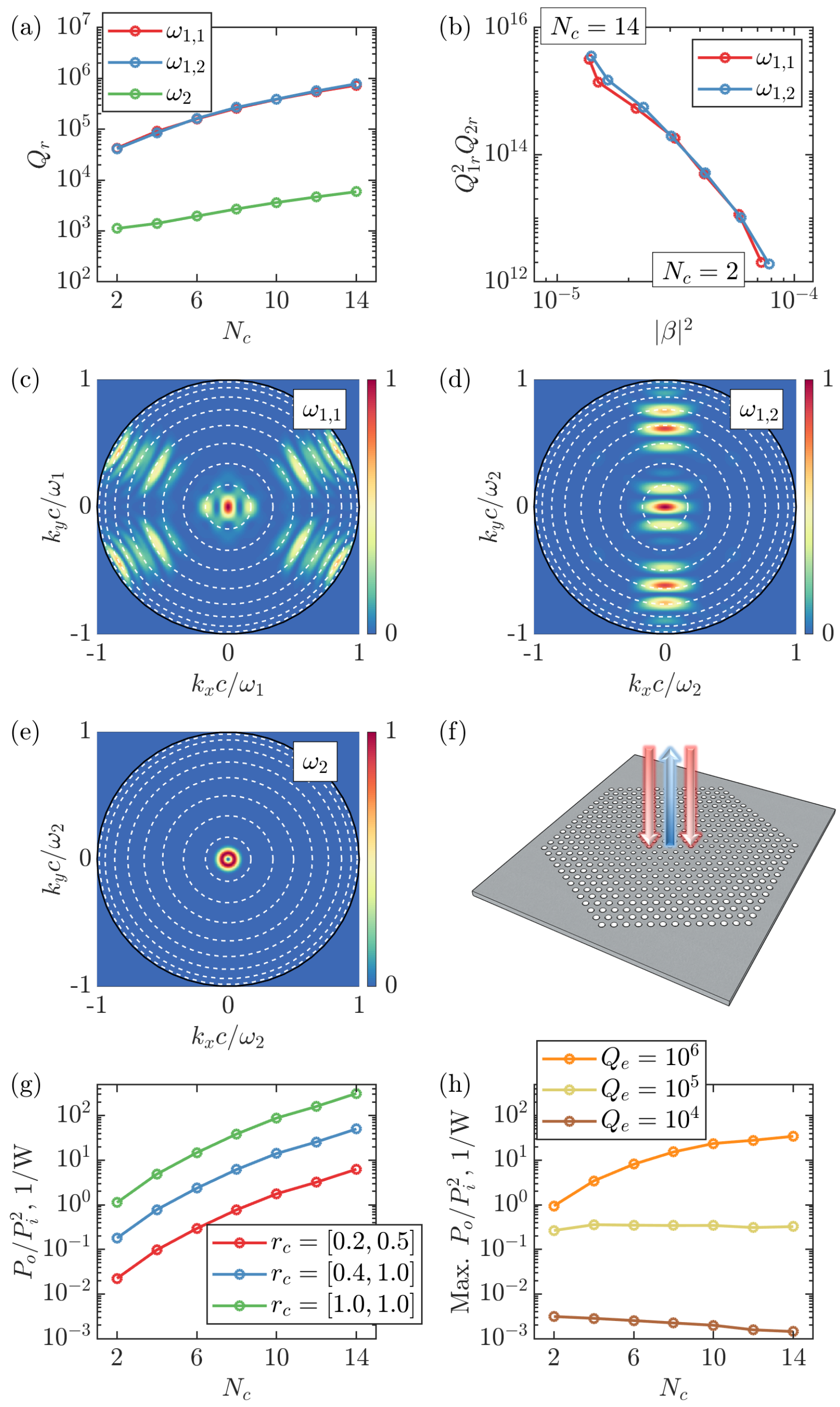}%
 \caption{(a): Radiative quality factors of the two FH modes and of the SH mode vs. core region size $N_c$. The rest of the heterostructure parameters are as in Fig. \ref{fig:het_cavity}. (b): The two figures of merit relevant to SH generation for the different core sizes $N_c$ shown in (a). (c)-(e): Polar far-field emission profiles in the upper half space for the two FH modes and the SH mode, respectively (cf. Fig. \ref{fig:het_cavity}). Dashed white lines show emission angles in steps of $10^\circ$. (f): Schematic of a proposed experimental setup for SH generation with pump excitation and signal collection from the vertical direction. (g): SH conversion efficiency in the limit of undepleted pump vs. $N_c$ for three different ratios $r_c$ of the overlap between the input/output channels and the heterostructure far field at the FH and SH frequencies, assuming no extrinsic losses ($Q_e \rightarrow \infty$). (h): Maximum attainable efficiency vs. core size for three different values of the extrinsic $Q_e$.}
\label{fig:q_beta}
\end{figure} 

\section{Cavity modes and figures of merit}

On the basis of these regular PhC modes, we can now introduce the cavity modes through the heterostructure design schematically shown in Fig. \ref{fig:phc_def}(b). We define three hexagonal regions: core, transition, and outer, with radii $r_c, r_t, r_o$, respectively. The size of the regions is defined by the number of hexagonal `layers' of holes in each of them $(N_c, N_t, N_o)$, or, alternatively, through the side-length of the hexagons, as explained in Fig. \ref{fig:phc_def}(b). Because the two bands in Fig. \ref{fig:phc_def}(c)-(d) have a \textit{minimum} at the band-edge frequency, the core region of the heterostructure is surrounded by holes with a \textit{larger} hole radius. In the cavity design that we study, we set $(r_c, r_t, r_0) = (0.22a, 0.23a, 0.25a)$ (see Supplement 1; also note that in the schematic of Fig. \ref{fig:phc_def}(b), the difference of the radii is larger for visualization purpose). The dotted lines in Fig. \ref{fig:phc_def}(c)-(d) thus show the bands for the regular PhC  in the outer region of the cavity, which lie, as needed, at a higher frequency than the core-region (solid) bands.

To simulate the electromagnetic modes of the heterostructure, we use Lumerical FDTD solutions -- a commercial finite-difference time-domain solver (see Supplement 1). These modes are shown in Fig. \ref{fig:het_cavity}(a)-(c) for $(N_c, N_t, N_o) = (6, 4, 14)$. In panel (a), we show the electric field of the SH mode at reduced frequency $\bar{\omega}_2 = \omega_2 a/(2\pi c) = 0.889$, corresponding to wavelength $\lambda_2 = 652$nm for a lattice constant $a = 580$nm. At the FH frequency, we find two degenerate modes at $\bar{\omega}_1 = 0.444$, shown in Fig. \ref{fig:het_cavity}(b)-(c). These profiles can be understood as heterostructure modes described by eq. (\ref{eqn:E_het}), with two envelope functions $f_{1, 1}(\mathbf{k})$ and $f_{1,2}(\mathbf{k})$. In our FDTD simulation, we also find higher-order modes at the FH frequency, but there is more than $5$nm spectral difference between those and the ones shown in Fig. \ref{fig:het_cavity}(c)-(d), even for the largest core size $N_c = 14$ studied below. At the SH frequency, the only mode that we find to be resonant in the core region and with the correct mode-matching symmetries for SHG is the one shown in Fig. \ref{fig:het_cavity}(a).

In Fig. \ref{fig:q_beta}(a), we show the radiative quality factors $Q_r$ of the three modes of the heterostructure as a function of the core region size $N_c$, while keeping $N_t = 4$, $N_o = 14$. As expected $Q_r$ increases with increasing $N_c$, and becomes particularly large for the FH modes -- approaching one million. The SH mode has a quality factor ranging from $1,100$ to $5,900$, which is also remarkably high given the fact that the mode is deep inside the light cone of the PhC, and that there is no SH photonic band gap. In Fig. \ref{fig:q_beta}(b), we plot the two important figures of merit for nonlinear second-harmonic generation: the product $Q_{1r}^2 Q_{2r}$, as well as a dimensionless version $\bar{\beta}$ of the nonlinear overlap factor of eq. (\ref{eqn:beta}) as defined in Ref. \cite{Lin2016}:
\begin{equation}
\bar{\beta} = \frac{\lambda_1^{3/2} \int \mathrm{d} \mathbf{r} \bar{\epsilon}(\mathbf{r}) \left(E^2_{1x} E^*_{2z} + E^2_{1y} E^*_{2z} \right)}{\left(\int \mathrm{d}\mathbf{r} \epsilon_1(\mathbf{r}) |\mathbf{E}_1|^2 \right)\left(\int \mathrm{d}\mathbf{r} \epsilon_2(\mathbf{r}) |\mathbf{E}_2|^2 \right)^{1/2}},
\label{eqn:beta_bar}
\end{equation}
where $\lambda_1$ is the free-space wavelength at the FH frequency, and $\bar{\epsilon}(\mathbf{r})$ is defined to be 1 in the nonlinear dielectric, and 0 outside. This dimensionless overlap factor allows to compare this figure of merit independently of the magnitude of the $\chi^{(2)}$ values and of the FH wavelength. We note that both $Q_r$ and the $\bar{\beta}$ associated to the two modes $\omega_{1, 1}$ and $\omega_{1, 2}$ are expected to be the same due to symmetry, and the slight difference is likely due to the FDTD simulation mesh and boundary conditions, which break the hexagonal symmetry. Furthermore, note that with larger $N_c$, the cavity becomes multimode with a decreasing frequency separation between the modes. These additional modes correspond to higher-order envelope functions $f_{n, p}(\mathbf{k})$ in eq. (\ref{eqn:E_het}). 

Finally, we consider a realistic experimental realization of second harmonic generation in our cavity, and estimate the conversion efficiency in the undepleted pump limit. For this, we will need to estimate the coupling coefficients of the radiative channels that will be used for pump in-coupling and SH signal out-coupling. Thus, in Fig. \ref{fig:q_beta}(c)-(d) we show the far-field emission profiles of the FH mode Fig. \ref{fig:het_cavity}(b) and the SH mode of Fig. \ref{fig:het_cavity}(a), respectively. In Supplement 1, we also show these profiles for a range of core size values for all three modes. These plots illustrate one additional advantage of our design in terms of second harmonic generation: namely, all emission profiles have a strong lobe within a small angle around the vertical direction. This holds particularly true for the SH mode, because its envelope function $f_{2}(\mathbf{k})$ is centered around $\Gamma$. Note that the emission at exactly zero angle is vanishing because of the BIC at $\Gamma$. Importantly, however, all of the generated second harmonic signal can be collected with a small numerical aperture from above the cavity, which is a significant improvement to previous studies using a photonic crystal platform. Furthermore, the strong lobe of the FH mode at zero emission angle facilitates free-space in-coupling of the pump radiation. We note that the vertical emission at the FH frequency is generally not guaranteed, but it is nevertheless observed for our particular design. In general, this can also be induced (or strengthened) through cavity modifications that specifically introduce $\mathbf{k} = 0$ components in the Fourier transform of the cavity modes \cite{Tran2009, Portalupi2010}.  

These considerations imply that efficient in- and out-coupling to the cavity can be achieved by exciting and collecting within a small angle around the vertical direction (schematically shown in Fig. \ref{fig:q_beta}(f)). To study the frequency conversion efficiency in such an experimental setup, we consider in-and out-coupling through part of the far field, assuming that a fraction of the radiative quality factor of the FH mode is due to coupling to the input channel. We define this fraction $r_{1c} = Q_{1r}/Q_{1c}$, where $Q_1^c$ is due to coupling to the pump mode and $Q_1^r$ is the radiative $Q_r$ of Fig. \ref{fig:q_beta}(a). Similarly, we write $r_{2c} = Q_{2r}/Q_{2c}$, where $Q_2^{c}$ is associated to the far-field modes from which the SH signal is collected. Furthermore, we include a phenomenological extrinsic quality factor $Q_e$, accounting for effects like linear loss or disorder-induced scattering, and write the total $Q_t$ of the modes as $Q_{it}^{-1} = Q_{ir}^{-1} + Q_{ie}^{-1}$ for $i = 1, 2$. In the limit of undepleted pump power $P_i$, the conversion efficiency $P_o/P_i^2$ for the SH output power $P_o$ can then be written as (see Refs. \cite{Rodriguez2007, Lin2016} and Supplement 1):
\begin{align}
\frac{P_o}{P_i^2} = \frac{8}{\omega_1}\left(\frac{\chi^{(2)}_{\mathrm{eff}}}{\sqrt{\epsilon_0 \lambda_1}}\right)^2 |\bar{\beta}|^2 \frac{Q_{1t}^4}{Q_{1r}^2} \frac{Q_{2t}^2}{Q_{2r}} r_{1c}^2 r_{2c},
\label{eqn:Pout1}
\end{align}
where $\epsilon_0$ is the permittivity of free space and $\chi^{(2)}_{\mathrm{eff}}$ is determined by the magnitude of the relevant elements in the susceptibility tensor. The conversion efficiency is maximized when there are no extrinsic losses, $Q_{ie} \rightarrow \infty$, and when the two ratios $r_{ic}$ go to unity, which happens when the pump mode completely overlaps with the far-field radiation pattern at the FH, and all the radiated SH signal is collected.

In Fig. \ref{fig:q_beta}(g), we plot the conversion efficiency vs. the core size $N_c$, assuming no extrinsic losses, for three different values of $r_{ic}$, reflecting three different scenarios. In the conservative case (red), $r_c = [r_{1, c}, r_{2, c}] = [0.2, 0.5]$, we assume excitation and collection from only one side of the cavity, and only a partial overlap of the pump mode with the far field pattern. In the optimized case (blue) $r_c = [0.4, 1.0]$, we still assume pumping from one side only, but with a better in-coupling of the pump mode, which can be achieved through far-field engineering. Furthermore, we assume that the SH signal is collected from both sides of the slab for optimal collection efficiency. Finally, the best-case scenario (green) assumes perfect in- and out-coupling at both frequencies, which would require either pumping the cavity from both sides or introducing a reflector on one side, and a perfect far-field overlap in either case. Notably, the values for the conversion efficiency exceed 100$\%/W$ even in the conservative case, and go beyond 10,000$\%/W$ in the best-case scenario and $N_c > 10$. 

In Fig. \ref{fig:q_beta}(h), we show the effect of extrinsic loss on the maximum achievable conversion efficiency for three different values of $Q_e$. This assumes $r_c = [1, 1]$, as well as some far-field engineering such that $Q_{1r}$ matches $Q_e$ (for details see Supplement 1). One notable effect of the extrinsic losses is to set the optimal core size of the cavity. Specifically, while the nonlinear overlap factor decreases with $N_c$ because of the increasing cavity mode volume (Fig. \ref{fig:q_beta}(b)), in the absence of external losses this is more than compensated by the increasing $Q_{1r}^2 Q_{2r}$ product, such that the efficiency in Fig. \ref{fig:q_beta}(g) is monotonically increasing with $N_c$. This is also the case for $Q_e = 10^6$ in Fig. \ref{fig:q_beta}(h). However, this changes for lower extrinsic quality factors, which limit the maximum total $Q_t$. Thus, the efficiency peaks at $N_c = 4$ for $Q_e = 10^5$, and $N_c = 2$ for $Q_e = 10^4$. We note that a total (i.e. loaded) quality factor of $44,000$ has already been experimentally measured for a GaN PhC cavity \cite{Mohamed2017}, suggesting that $Q_e$ approaching $10^5$ should be realistic in state-of-the-art systems.

\section{Conclusion}

We conclude by discussing the merits of the cavity presented here in comparison to previous works. The quantity $\bar{\beta}$ shown in Fig. \ref{fig:q_beta}(b) is defined in a way that is particularly useful for comparing to designs in other materials and wavelengths. A comparison to the devices summarized in Ref. \cite{Lin2016} shows that the cavity presented here is orders of magnitude better than previous PhC designs either in the $Q$-product, or in the nonlinear overlap $\bar{\beta}$, or in both. In fact, the only designs that outperform our cavity in terms of $\bar{\beta}$ are the topology-optimized micropillars \cite{Lin2016} and microrings \cite{Lin2017}, which are, however, more challenging to fabricate in practice. Furthermore, the topology-optimized micropillars \cite{Lin2016} exhibit a significantly lower $Q_1^2 Q_2$ product and therefore a much lower conversion efficiency as per eq. (\ref{eqn:Pout1}) and FOM1 defined in \cite{Lin2016}, while the multi-ring designs of Ref. \cite{Lin2017} cannot be side-coupled to an input/output channel, and also offer little engineering possibility for out-of-plane emission. Alternatively, large microring cavities could support modes with a $Q_1^2 Q_2$ product exceeding the one of our design, but at the expense of a much larger footprint \cite{Bruch2018}. In short, our design combines a high nonlinear overlap with a high $Q$ product in a very compact cavity, with the additional advantage of strong vertical emission at both frequencies. 

More broadly, we have outlined a general recipe for designing high-$Q$, doubly-resonant photonic crystal cavity modes with a high nonlinear overlap that can be applied to various materials and setups. This could be useful to further improve the design presented here, as the fact that a lot of the field of both modes is in air indicates that there is enormous room for improvement of the overlap factor. Thus, for example by finding another suitable pair of bands according to our recipe, by exploring PhCs formed by dielectric rods in air, or by infiltrating the PhC holes with a nonlinear material, the nonlinear overlap could be further increased to record-high values. Finally, we further notice that the doubly resonant cavity design proposed here naturally allows for strong field enhancement in the air holes, both at FH and SH. This unique characteristic might be exploited, e.g., for strongly enhanced nonlinear sensing of functionalized surfaces \cite{Estephan2008, Estephan2010}. All in all, we believe that the generality of our approach coupled with the well-established fabrication methods for photonic crystal slabs could enable extremely compact devices for second-order nonlinear conversion with unprecedented efficiency.

\section{Acknowledgements}

This work was supported by a MURI project of the Air Force Office of Scientific Researches (FA9550-17-1-0002) and the Swiss National Science Foundation through Project N\textsuperscript{\underline{o}} P300P2\_177721. DG acknowledges enlightening discussions with M. Liscidini and M. Galli, and financial support from the EU H2020 QuantERA ERA-NET Cofund in Quantum Technologies project CUSPIDOR, cofunded by the Italian Ministry of Education and Research (MIUR).

\bibliography{double}

\end{document}